# Vortex-induced vibrations and wake induced oscillations using wake oscillator model
# Comparison on 2D response with experiments

David Cébron, Benoît Gaurier, Grégory Germain

*IFREMER, Hydrodynamics and Metocean, 62321 Boulogne-sur-Mer, France*
*Corresponding author:G.GERMAIN: ggermain@ifremer.fr*

Mooring and flow lines involved in offshore systems for oil production are submitted to various solicitations. Among them the effects of current are dominating. Vortex-Induced Vibrations (VIV) and Wake-Induced Oscillations (WIO) on closely spaced marine risers may lead to fatigue, clashes and structural failures. Extended studies have been conducted to describe and explain them for spring mounted uniform cylinders in translation perpendicularly to their main axis [1]. In the case of a pivoted cylinder with uniform diameter [2] a similar response is observed with some variations on the reduced velocity interval and maximum response. However, complementary studies are necessary to improve our understanding of the hydrodynamic loads acting on risers placed in the wake of an upstream one.

From experiments on model scaled tests with real configurations for dual riser interaction in uniform and steady current performed in the IFREMER circulation water tank, we have obtained information on the behaviour of two risers exposed to steady current and excited by VIV and WIO. These tests give a lot of information on how fluid interaction between two cylinders of equal diameter in tandem configuration can significantly modify their structural response in term of amplitude and frequency, compared to that of a single one [3], [4]. Both in-line and cross-flow responses have been studied and presented as functions of the reduced velocity. Results demonstrate that wake effects can be relatively strong. If in almost all the cases the upstream cylinder responds like an isolated single one whereas the vortex shedding and synchronization of the downstream cylinder can be strongly affected by the wake of the upstream one. We have shown that the excitation of cylinders may produce large motions, sometimes enough to bring them into contact at relatively high reduced velocities (Vr > 19).

Those results are here used to validate a 2D phenomenological model of the near wake based on Van Der Pol wake oscillators [5], [6] which is the first one which describes the 2D motion of the cylinder in its transversal plan. This simplified model of the wake dynamics is first validated on a single cylinder for which we consider the 2D response. The coupled fluid-structure dynamical model is given by:

$$\begin{cases}(m+C_m\rho V)\cdot\ddot{x}+\lambda\cdot\dot{x}+k\cdot(x-x_0)=\frac{1}{2}\rho S\left[(C_D+C_{Dk})(U_x-\dot{x})-C_{Lk}(U_y-\dot{y})\right]\sqrt{(\dot{x}-U_x)^2+(\dot{y}-U_y)^2}+\rho V\underbrace{(1+C_m)}_{C_M}\dot{U}_x+\frac{1}{2}\rho SC_{D_{trans}}\sqrt{U_x^2+U_y^2}\cdot U_x \\ (m+C_m\rho V)\cdot\ddot{y}+\lambda\cdot\dot{y}+k\cdot(y-y_0)=\frac{1}{2}\rho S\left[(C_D+C_{Dk})(U_y-\dot{y})+C_{Lk}(U_x-\dot{x})\right]\sqrt{(\dot{x}-U_x)^2+(\dot{y}-U_y)^2}+\rho V\underbrace{(1+C_m)}_{C_M}\dot{U}_y+\frac{1}{2}\rho SC_{D_{trans}}\sqrt{U_x^2+U_y^2}\cdot U_y \\ \ddot{C}_{Lk}+\varepsilon_{Lk}\omega_{C_L}\left(\left(\frac{C_{Lk}}{C_{Lk_0}/2}\right)^2-1\right)\dot{C}_{Lk}+\omega_{C_L}^2\cdot C_{Lk}=A_{Lk}(\ddot{y}-\dot{U}_y) \\ \ddot{C}_{Dk}+\varepsilon_{Dk}\cdot 2\omega_{C_L}\left(\left(\frac{C_{Dk}}{C_{Dk_0}/2}\right)^2-1\right)\dot{C}_{Dk}+(2\omega_{C_L})^2\cdot C_{Dk}=A_{Dk}(\ddot{x}-\dot{U}_y)\end{cases}$$

where the parameters *m, k, $x_0$, $y_0$, λ, S, V, ρ, $U_x$, $U_y$* characterised the studied system and the parameters $C_{Dk}$, $C_{Lk}$, $A_{Lk}$, $A_{Dk}$, $\varepsilon_{Lk}$, $\varepsilon_{Dk}$, $\omega_{Cl}$ fix the dynamic of the system under closed form relations (will be given in the future paper). Those equations are enough general to predict precisely a drop in still water without modification and extended to two cylinders in tandem [9].

In the case of a single cylinder, experimental and numerical results are in relatively good agreement, like shown on the figures below, where the rms and mean amplitude response of the system are presented as a function of the reduced velocity. The lock-in region is here well evaluated: the maximum displacement amplitude of the transverse response and the mean in-line displacement at lock-in are in good accordance. The gap between the in-line RMS amplitude responses may come from the lack of data in the litterature about the RMS drag coefficient and its growth factors because we have to use mean values. On the contrary, the RMS lift coefficient is well described in the model and takes into account the effect of the length of correlation along the cylinder on the total lift [7], which is completely new in this kind of model.

A recent paper [8] provides new measurements and so allows us to extend the validation of the proposed model and to improve its accurracy, especially at lock-in, with a better description of the RMS drag coefficient (in development).

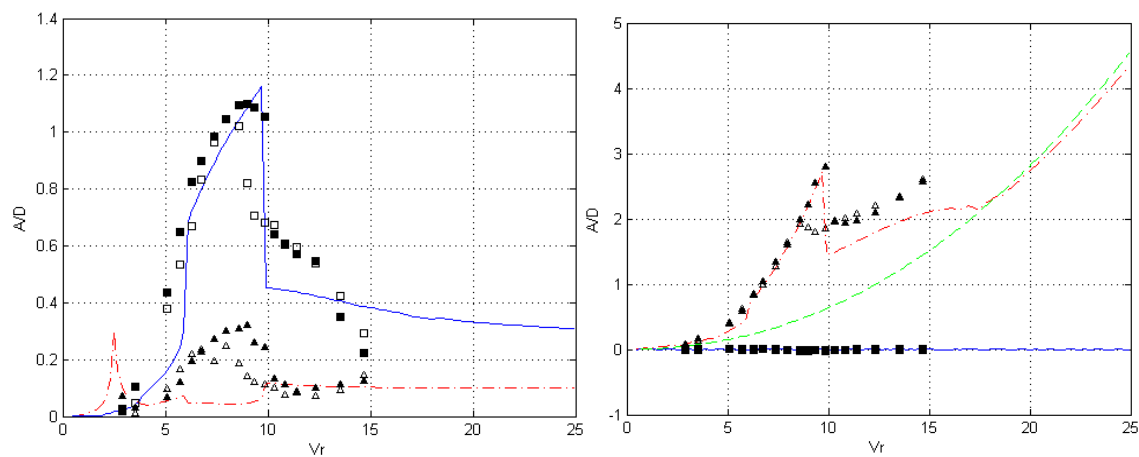

Rms and mean amplitude response as a function of reduced velocity (Vr)
□, ■ and — : transverse motions ; ∆, ▲ and -- : in-line motions   (symbols for experimental results)

Those first results show that the proposed model can be used as a simple computation tool in the prediction of 2D VIV effects. Its extension to model wake effects of two cylinders in tandem arrangement is in development and will be compared to experimental results. This model completes previous studies [5] from which a 3D phenomenological model for long cables could be obtained in a near future.